# Microservices Migration in Industry: Intentions, Strategies, and Challenges

Jonas Fritzsch*†, Justus Bogner†*, Stefan Wagner*, Alfred Zimmermann†
* University of Stuttgart, Germany
† University of Applied Sciences Reutlingen, Germany
{jonas.fritzsch, stefan.wagner}@informatik.uni-stuttgart.de
{justus.bogner, alfred.zimmermann}@reutlingen-university.de

*Abstract*—To remain competitive in a fast changing environment, many companies started to migrate their legacy applications towards a Microservices architecture. Such extensive migration processes require careful planning and consideration of implications and challenges likewise. In this regard, hands-on experiences from industry practice are still rare. To fill this gap in scientific literature, we contribute a qualitative study on intentions, strategies, and challenges in the context of migrations to Microservices. We investigated the migration process of 14 systems across different domains and sizes by conducting 16 in-depth interviews with software professionals from 10 companies. We present a separate description of each case and summarize the most important findings. As primary migration drivers, maintainability and scalability were identified. Due to the high complexity of their legacy systems, most companies preferred a rewrite using current technologies over splitting up existing code bases. This was often caused by the absence of a suitable decomposition approach. As such, finding the right service cut was a major technical challenge, next to building the necessary expertise with new technologies. Organizational challenges were especially related to large, traditional companies that simultaneously established agile processes. Initiating a mindset change and ensuring smooth collaboration between teams were crucial for them. Future research on the evolution of software systems will in particular profit from the individual cases presented.

## I. INTRODUCTION

Monolithic systems that have grown over years or even decades can become large and complex and even fossilize in later stages [1], making them practically unmaintainable. In the course of a modernization, several companies migrate their legacy systems towards a Microservices architecture, aiming for better maintainability, shorter release cycles, scalability, cloud-readiness, or high availability. Refactoring a system's architecture to Microservices has many implications, depending on the company culture and established processes. To efficiently develop and operate Microservices, a combined flow of development and operation with a high degree of automation, also referred to as DevOps [2], is beneficial. For this purpose, cross-functional teams are built around business capabilities and follow an agile process model. The importance of a well planned migration arises from its high cost, long duration, and involved organizational restructurings next to the architectural refactoring itself.

Several publications discuss benefits and challenges of adopting Microservices [3], [4], address design aspects [5] or patterns [6]. Studies also state that industrial state-of-practice has already reached some degree of maturity, while academia is still at an early stage [7], [8]. According to Vural et al., this equally applies to empirical studies that provide insights into the current state-of-practice [9]. Moreover, Jamshidi et al. note that recently published papers have had "*little if any impact on microservice practice*" [10], which according to the authors might be caused by limited access to industry-scale applications. A recent mapping study again confirms a strong industry interest in migrating legacy systems [11].

To address this gap and to provide additional empirical and industry-focused research, we conducted 16 in-depth interviews with software professionals based in Germany from 10 different companies. We talked with them about 14 Microservices-based systems, nine being in the course of a migration, three having migrated recently and one greenfield development. Our goal was to identify intentions and strategies for the migrations as well as challenges that companies were faced with. By focusing on the individual cases separately, we aim to also provide rationales and successfully applied practices.

## II. RELATED WORK

Some case studies report on Microservices migrations in a single company [12] or domain [4]. Balalaie et al. extract common patterns from three industrial migration projects [13], while Di Francesco et al. aimed to achieve more generalizable insights on migration activities and challenges by consulting 18 practitioners [14]. The results focus on activities and faced challenges per migration phase. Taibi et al. went one step further by deriving a migration process framework out of 21 interviews with experienced practitioners, who already migrated their monoliths to Microservices [15]. They identified maintainability and scalability as main drivers, while decomposing the monolith and splitting the data remained major technical challenges. Similar aspects are covered by Knoche et al. [16]. In their survey among 71 participants, the authors differentiate their results by industry sector and identify scalability as the main driver for early adopters, while traditional businesses rather aim for maintainability in the long term. Likewise, the survey by Ghofrani and Lübke [17] focuses on challenges during a migration.

In summary, two of the studies [14], [15] interviewed groups of similar size, but provide aggregated results only. In contrast, our study presents additional insights and experiences from practitioners through a discussion of each system. We aim to substantiate and enhance the aforementioned findings by a holistic view of 14 industrial Microservices migration cases, that involves organizational aspects as well.

## III. RESEARCH METHOD

We followed a qualitative approach by conducting semi-structured interviews [18], [19]. Qualitative methods aim at understanding phenomena within their real life context [18], directly deal with existing complexity and are useful for addressing the *how* question [20]. This seems appropriate for intentions, strategies, and challenges of a migration scenario, as they are tightly linked to their context. Furthermore, interviews provide the flexibility to probe for further details on specific aspects, enabling us to focus on participants' rationales. Despite the significant effort for preparing, planning, recruiting, and conducting the interviews, the method was best-suited for our research goal. We followed the guidelines proposed by Runeson and Höst [21], who advocate the execution of five steps: 1) Case study design, 2) Preparation for data collection, 3) Collecting evidence, 4) Analysis of collected data, and 5) Reporting. In the following, we outline our approach.

**Study design:** Our overarching research objective is to analyze the migration process from monolithic architectures to Microservices on the basis of real-world systems in industry practice. To get a holistic view, we aim to identify intentions, applied strategies, and challenges, which can be related and therefore should be examined coherently. To increase the transparency of our findings, we describe the main aspects for each case individually. This objective is framed by the following three research questions:

**RQ1:** What are intentions for migrating existing systems to Microservices?
**RQ2:** Which Microservices migration strategies and decomposition approaches do companies apply?
**RQ3:** What are the major technical and organizational challenges during a Microservices migration?

**Preparation and evidence collection:** An interview preamble [21] was created and sent to the participants beforehand to make them familiar with the study. It explained the interview process and topic, but also aimed to maintain ethical standards by ensuring confidentiality and asking for consent to create audio recordings. To ensure comparability of the interviews, an interview guide for the moderators was set up. It contained the interview questions along with remarks and provided a basic structure.

The first two authors conducted the interviews in the fourth quarter of 2018. Of the 16 individual interviews, six were conducted in person and 10 via remote communication software with screen sharing. Before the recording started, participants were again asked for their consent. Anonymity was ensured, as well as the deletion of the recordings after transcription. During the interviews of ~45 to ~75 minutes, the prepared guide was loosely followed and adapted based on participants' reactions. Shortly after each interview, the recording was transcribed and sent to the participant for review and final approval. Hence, participants were able to remove sensitive information, change incorrect statements, or improve wording. The subsequent analysis built upon the returned transcripts.

**Data Analysis:** We used the software toolkit MAXQDA for qualitative data analysis to perform coding of each transcript. From a methodological viewpoint, coding involves the formation of categories (codes) and assignment of fitting text parts to these categories. In a first run, all statements answering a specific interview question were coded. This improved navigation in the transcripts and helped to easily extract e.g. demographic data. Afterwards, a set of codes for the research questions was created. The main categories were defined as *Intentions*, *Strategies*, and *Challenges*. In the subsequent second run, all transcripts were coded using this system. During the course of it, new sub-codes emerged and existing ones were renamed, split, or merged as we acquired a better understanding of the cases. When a new aspect was discovered and a code created respectively, all already encoded interviews were reviewed to that effect. This approach follows the constant comparison method based on Grounded Theory [18]. For interpretation of the coded transcripts, different strategies were followed. The encoded parts were reread and statements were connected, since different formulations often had the same meaning. Based on the coding system, tabulation [21] was used to obtain an overview of the data. These tables were used to identify results and draw conclusions. To ensure traceability, we provide our interview guide, code system, and code frequencies in an anonymous online repository[1].

## IV. INTERVIEW CASE DESCRIPTIONS

The participants were software professionals in technical roles (developer or architect) with significant professional experience (minimum 5 years) and solid knowledge of service orientation. We required participation in the implementation of a Microservices-based system, primarily in a migration context. Participants were recruited via inquiries to several companies in addition to existing contacts within the research group. All participants were based in Germany, even though some of the companies were active in several European countries or globally. Details of the participating companies (C1–C10) of various sizes and domains are shown in Table I. Most of them were software & IT services companies that develop systems for external clients. For the companies from other domains, system ownership was always internal. Altogether we investigated 14 systems (S1–S14, see Table II). We summarized the following case descriptions from participants' explanations, based on the system in place.

---

[1] https://figshare.com/s/d7cb071a527503435a9d

TABLE I
COMPANY AND PARTICIPANT DEMOGRAPHICS

| Company ID | Company Domain | # of Employees | Participant ID | Participant Role | Years of Experience | System ID |
|---|---|---|---|---|---|---|
| C1 | Financial Services | 1 - 25 | P1 | Developer | 6 | S1 |
| C2 | Software & IT Services | >100,000 | P2 | Lead Architect | 30 | S2 |
| | | | P3 | Architect | 24 | S3 |
| | | | P4 | Architect | 30 | S4 |
| C3 | Software & IT Services | 26 - 100 | P5 | Architect | 20 | S5 |
| | | | P6 | Lead Developer | 8 | |
| C4 | Software & IT Services | 101 - 1,000 | P7 | Architect | 9 | S6 |
| | | | P8 | Architect | 17 | S7 |
| C5 | Software & IT Services | >100,000 | P9 | Lead Developer | 7 | S8 |
| C6 | Tourism & Travel | 1,001 - 5,000 | P10 | Developer | 9 | S9 |
| | | | P11 | Data Engineer | 7 | |
| | | | P12 | Architect | 12 | S10 |
| C7 | Logistics & Public Transport | 101 - 1,000 | P13 | DevOps Engineer | 5 | S11 |
| C8 | Retail | 5,001 - 10,000 | P14 | Lead Architect | 9 | S12 |
| C9 | Software & IT Services | 101 - 1,000 | P15 | Architect | 18 | S13 |
| C10 | Retail | 1,001 - 5,000 | P16 | Architect | 22 | S14 |

**C1-S1, Derivatives Management System:** C1, a small financial services company, offers a system for the management and search of derivatives as Software as a Service (SaaS). The single-tenant solution evolved to a complex monolith over time. According to P1, it was hard to make changes in the code due to the high complexity and internal dependencies. The decreasing maintainability made it difficult to release new functionality in a timely manner. Another migration driver was the planned scaling of organizational units. The small team decided for a gradual migration without involving external resources. The team successively extracted functionality from the PHP monolith, starting with parts that did not necessarily need to run inside the central instance. This resulted in several smaller services grouping around the core service. No structured decomposition approach was followed. Challenges arose around finding a solution for service discovery, which was achieved by static configuration in the beginning. Eventually, the team found a reasonable solution with Eureka (Spring Cloud). Another challenge was to achieve a certain degree of fault tolerance by "*reacting in a graceful way*" (P1) to failing or missing services. As an intermediate result of the ongoing migration, P1 stated a significant reduction of time to market for new features.

**C2-S2, Freeway Toll Management System:** The first of three systems developed by the large software consulting enterprise C2 was a management and payment system for freeway tolls. The initial decentralized solution was replaced by a central Microservices-based system with extended functionality. According to P2, the system suffered from degraded maintainability and didn't meet the current requirements anymore, like the increased amount of transmitted sensitive data and compliance with high security standards arising thereby. A major technical challenge was to find the right service cut. Here, P2 advocated for a greenfield-like approach to develop the new system: "*Most productive and successful Microservices projects that I know are greenfield developments.*" He was afraid of the complex monolith and the necessary effort to replace transactions by eventual consistency. As well, he was skeptical towards concepts like Domain-Driven Design in this regard and convinced that there is no way to automate such a task. Other major challenges were the integration of services and communication with external parties, as well as test automation. The need to migrate the three core services within a short time resulted in slightly oversized development teams, which had a negative impact on the team's efficiency. As a result of the completed migration, the new system fulfilled all requirements. For assessing its maintainability, it would still be too early.

**C2-S3, Automotive Problem Management System:** The second system from C2 was an automotive management system for the categorization and analysis of problems. The very traditional Java monolith with JSF frontend had many flaws. Users started building their own workarounds using e.g. Excel. To increase maintainability, the customer decided to rewrite and extend the solution. According to P3, further expectations for the new architecture were scalability and the facilitation of a multi-vendor strategy. However, the customer's decision for Microservices would not have been taken very rationally.

The migration was started by rebuilding a user-facing component using current technologies and integrating it with the legacy monolith. This component served as a foundation for successively building a service-based solution along the entire business workflow, with each bounded context forming a service. There was no particular method followed to determine the service cut or achieve appropriate granularity. P3 also notes that the decomposition partially fell short in terms of loose coupling. A change to one service sometimes involved adapting two or three others. Another challenge was keeping both systems in sync during the migration phase. Here, the initially rebuilt landing page was used to route the traffic between both systems. P3 emphasized the vast amount of new technologies that had to be mastered in a short time, which implied learning efforts on the way. The customer originally followed a Waterfall model, which was gradually replaced by agile processes based on Scrum. The ongoing transition turned out to be a difficult process in this regard. Nonetheless, the agile principles were perceived positively by the customer. P3 found it interesting to see how people were more motivated and sometimes showed themselves from a different side.

TABLE II
SYSTEM OVERVIEW AND ORGANIZATION

| ID | Purpose | Inception | Timeframe of Migration in years | # of Services | # of People involved | Team Size | Process Model |
|---|---|---|---|---|---|---|---|
| S1 | Derivatives mgmt. (banking) | Rewrite | 1.75 (ongoing) | 9 | 7 | 7 | Scrum |
| S2 | Freeway toll management system | Rewrite & Extension | 1.5 to 2 | 10 | 10 (only devs) | 5-10 (up to 40) | Individual (based on Scrum) |
| S3 | Automotive problem management system | Rewrite & Extension | 2 to 3 (ongoing) | 10 | 50 | 7-9 | Scrum (from Waterfall) |
| S4 | Public transport sales system | Rewrite & Extension | 2 (ongoing, exp 4) | ∼100 | ∼300 | 6-10 | Scrum, SAFe (from Waterfall) |
| S5 | Business analytics data integration system | Greenfield | 1.5 to 2 | 6 | 7 | 7-9 | Individual (based on Scrum, Kanban) |
| S6 | Automotive configuration management system | Rewrite | 0.5 (ongoing, exp 3) | 60 | 20 (w/o cust.) | 4 | Scrum (from Individual) |
| S7 | Retail online shop | Replace COTS | 2.5 (ongoing) | ∼250 | ∼200 | 6-8 | Scrum, Kanban |
| S8 | IT service monitoring platform | Cont. Evolution | 2 (ongoing, exp 3) | 9 | 15 | 6-10 | custom |
| S9 | Hotel search engine | Cont. Evolution | 1 to 1.5 (ongoing) | ∼10 | ∼50 | 3-6 | Scrum |
| S10 | Hotel management suite | Rewrite & Extension | 0.5 to 1 (ongoing, exp 2) | 20 | 50 | 1-5 | Scrum, Kanban |
| S11 | Public transport mgmt. suite | Cont. Evolution | 2 to 3 (ongoing) | 10 | ∼175 | 5-8 | Scrum |
| S12 | Retail online shop | Replace COTS | 1.5 | ∼45 | ∼85 | 6-10 | Scrum, Kanban |
| S13 | Automotive end-user service management | Rewrite & Extension | - (ongoing) | 7 | 30 | 5-7 | Scrum |
| S14 | Retail online shop | Replace COTS | 2.5 | ∼175 | ∼350 | 6-10 | Scrum, Kanban |

**C2-S4, Public Transport Sales System:** The last system from C2 was a sales system for public transportation. The legacy system was a complex of four monoliths that grew over more almost 40 years and provided an enormous set of functionality. The monoliths were not well integrated, which often required duplication of development efforts. Microservices seemed to be the fitting architecture to accommodate the system in manageable chunks. P4 described the chosen approach as something between green- and brownfield. In fact, the system was completely rebuilt, but adjacent systems remained unchanged. The service cutting was guided by the functional areas of the originating system, resulting in ∼100 Microservices structured in 15 domains. Hereby, Domain-Driven Design was used without involving additional tools.

The biggest challenge for this large-sized system was the alignment of activities for the 300 developers in 30 teams, e.g. maintaining a common backlog for functionality that affects multiple teams. Likewise, orchestrating the resulting Microservices was seen as challenging on the technical side. To mitigate the issues, C2 created a "*very large amount of guidelines and rules for service creation*" (P4). Still, the resulting service granularity was very different. The process model had been adapted from classic Waterfall to SAFe (Scaled Agile Framework) and Kanban in some teams, which was rolled out one year after the migration started. P4 observed an efficiency increase after a settling-in period of around one more year. According to P4, the decreased number of defects found via manual testing proved the positive impact of test automation by the implemented CI/CD pipelines.

**C3-S5, Business Analytics & Data Integration System:** The small software and consulting company C3 develops a business analytics system for big data that is focused on data integration and analysis via data mining or linguistic algorithms. We interviewed architect P5 and lead developer P6. The company decided to build the central backend in a greenfield approach based on Microservices. Intentions were building a manageable, maintainable and scalable platform which was seen as a mandatory capability in the *big data* context. The composition of six Microservices was designed by the lead architect.

Both interviewees saw building the necessary expertise as challenging for a small team. It would hardly be possible to have dedicated experts for all areas. P6 remarked that it required developers with the right attitude, who are eager acquire new knowledge even after work, if necessary. Still, this situation led to compromises in architectural design for the sake of simplicity, e.g. a single service accommodating the entire data. Another challenge was achieving fault tolerance among services, which evolved into "*lengthy philosophical and controversial discussions*" (P6) on how bullet-proof the solution needed to be. Furthermore, versioning and security aspects were brought up, whereas the latter ones were crucial for the product and got resolved in a time-devouring process. The team of C5 employs agile methods based on Scrum and Kanban, but got slightly off track due to the pressuring day-to-day business which sometimes "*needs to be managed in a traditional way*" (P5). Efforts to achieve pipeline automation were currently in progress, but hindered by the lack of experienced personnel and tasks of higher priority.

**C4-S6, Automotive Configuration Management System:** The software services company C4 was in the process of modernizing a very large monolithic configuration management system for an automotive customer. P7 was impressed by its sheer size, which likewise applied to the processes: "*A significant change takes ludicrous 6 months to go into production.*" The customer opted for a Microservices architecture for a number of reasons, including extensibility, time to market, scalability of components and teams, avoiding maintenance-downtime, and various new functional requirements.

According to P7, a break down of the monolith was not an option due to its huge size, business criticality, and included proprietary components. The only feasible option was moving the functionality to another platform. To deal with the peculiarities of the system and equally gain the trust of the customer's workforce, the architects chose a two-step approach. At first, all functionality based on read operations was moved. This tremendously reduced the load of the legacy system and would ease the more critical second step. P7 outlined the subsequent extraction of functionality as follows: 1) start at the outer layer with documented and external functionality, 2) gradually move to the inner structure and isolate, 3) coordinate with architect and business unit and 4) extract to a new service. The very comprehensive documentation of functionality was key for this approach. P7 was aware that the achieved cut will never fulfill all requirements, but the earned flexibility in merging/splitting later on would compensate for it. To avoid too fine-grained services, P7 followed the rule "*per microservice a 4-developers team and 3 months to production*".

A major technical challenge was providing 15 TB of data to all consumers in a timely manner and in this regard familiarizing stakeholders with the concept of eventual consistency. The customer also aimed to establish agile processes in conjunction with a mindset change towards a *fail fast* philosophy. Prototyping and rolling releases helped the team members to gain confidence and support the transition towards agile processes. Dissolving the strict separation between business and IT teams was another important step to gain momentum. After a "*bumpy start at day one*" (P7), the customer very much appreciated the tremendous improvements achieved so far, especially in time to market.

**C4-S7, Retail Online Shop:** The second system developed by C4 was an online shop for a large retailer. The existing commercial off-the-shelf (COTS) solution was incapable to fulfill the growing need for new functionality. Moreover, the customer aimed for scalability in terms of system users and development teams. Consequently, a new eCommerce platform based on a Microservices architecture should replaced the existing solution. For this, functionality was gradually re-implemented as services and new ones were created as required. So far, the nearly 200 involved people had created ∼250 services that were structured by domains and products. Since only the use cases were transferred from the existing solution, P8 described it as a greenfield approach. Integrating the various 3rd-party systems with the new architecture was seen as a challenge. P8 saw technical matters as always manageable, something which can be acquired with the help of external partners, if necessary. The customer had established agile processes before already and tended more towards practicing Kanban lately. Organizationally, he saw the main challenge in building a future for existing teams in the new environment and empowering them to deal with new technologies.

**C5-S8, IT Service Monitoring Platform:** The global IT services provider C5 developed an IT monitoring solution with an operations dashboard. Lead developer P9 described the initial architecture as a RESTful monolith that was currently migrated towards a more fine-grained architecture. Driver for breaking up the system was its quick growth over the last years. It had a negative impact on maintainability and increased the on-boarding time for new developers. Technically, the system had reached its resource limits, too. As a consequence, a first memory-intensive component was isolated and moved to a separate system. The approach was repeated for other components, where already isolated components had been extracted and partly split up further. P9 stated that the aspect of service re-usability was considered for splitting, but in general no systematic decomposition approach was followed. Over a time frame of almost two years, nine services had been extracted that way.

A major technical challenge was setting up the Kubernetes environment with its peculiarities in persistence, logging, and security aspects. P9 terms the applied process model as agile, without an underlying framework. The organization strives towards functional teams with a separation of development, operations, and support. This should help developers to concentrate on their core competencies across several services. Only external colleagues would contribute to single services exclusively, which compensates for the shortage on skilled personnel. In this regard, P9 described the knowledge transfer between all contributors as one of the biggest challenges to date.

**C6-S9, Hotel Search Engine:** Developer P10 and data engineer P11 both talked with us about S9, an online search platform for hotels. The decision for implementing Microservices resulted from a company-wide strategy. It involved aiming for cloud-readiness and scalability, to deal with the increasing and dynamically changing traffic caused by search queries to the system. Next to performance bottlenecks in conjunction with data storage, the monolithic PHP application suffered from a decreasing time to market for new features. In the course of implementing the company's strategy, management approved to rebuild the existing system. The overall strategy followed a gradual rebuild and extension of the entire functionality, based on the 12-factor-app principles [22]. According to P10, except for some special logic the entire system was newly built. P10 emphasized the achieved flexibility with regards to splitting and merging services and saw a tendency towards more fine granular services whereas data engineer P11 was skeptical about this direction and the increased traffic overhead. The switch from a central database to a stream-based architecture was seen as one of the challenges, next to mastering new technologies (e.g. gRPC) and finding best practices for building Microservices in general (e.g. deployment or version control). Running the legacy and the new system in parallel during the migration was seen as challenging due to the necessary synchronization.

Teams were realigned in the course of the migration and follow Scrum or Kanban process models. P10 emphasized a missing "*high level scope*" during the transition which resulted in frequent unforeseeable changes. In particular, realignments that led to discarding activities right before completion were frustrating for the teams. P11 expressed similar concerns and moreover was skeptical about politically motivated changes in general. Still, both interviewees positively rated the achieved efficiency and collaboration in their teams.

**C6-S10, Hotel Management Suite:** The second system from C6 was a management suite for hoteliers with several modular products. Intentions for migrating the 5-6 years old monolith were similar to S9, whereas P12 would add a tendency towards a modern and trending architectural style. The strategy used here was a functional decomposition with vertical isolation into 20 services following the Self-Contained Systems (SCS) paradigm. The existing code was transferred to the new system in small chunks. P12 made experiences with other Microservices-based systems already and stated that the decomposition approach always depended on the business case and could not be automated in any way. He was not afraid of such a challenge and made good experiences with learning from developers who built the original systems. However, he also stated that the creators of legacy applications were often unavailable and even if they were available, "*it can be luck or a challenge, depending on the guy*" (P12). He explicitly neglected Domain-Driven Design due to its perceived complexity, given the various other challenges that will arise in a migration. Teaching and convincing developers to think in new ways and "*to forget about foreign keys and relational databases*" (P12) was one of his tasks. Likewise, it was also important for him to prevent "*hipsterization*": "*Most people that do new stuff have a tendency to first choose the technology and then make a decision*" (P12).

**C7-S11, Public Transport Management Suite:** C7 developed a product suite for managing the whole process of public transportation (S11). DevOps engineer P13 described the ongoing modernization of the whole suite, which is accompanied by occasional team reorganizations and introduction of agile methods. The current modernization efforts of the nearly 30 years old ecosystem aimed to standardize and automate deployment processes and simultaneously achieving a more fine-grained architecture. The originally project-driven developed and highly customizable solution showed many symptoms of bad maintainability. As a very small team drove the modernization, only new functionality was implemented as separate services by using new technologies: "*Migration without feature development does not happen due to time constraints*" (P13). Goal of the architecture group was to let the monolith starve out in the long term, without needing to touch old code again. The integration of new services with the distributed legacy application proved difficult at times. In this regard, the increased testing effort was an issue. As the product was operated by partly "*IT-conservative*" customers, switching to containers was still not yet discussed. Building up the required expertise on a large scale in the established organization was also seen as a major challenge.

**C8-S12, Retail Online Shop:** Retailer C8 was operating a COTS online shop (SAP Hybris). The existing solution was limiting feature development and dynamic scaling. For the decision to develop a custom in-house eCommerce platform (versus updating Hybris), time-to-market was crucial. According to P14, this is an important aspect for online retailers who act in an extremely fluctuating market. While the existing solution remained in production, a greenfield Microservices project was started. As *shopping* would be a long established process, the domain cut was more or less predetermined. For that, the team relied on Domain-Driven Design and event storming techniques. Services were further divided into sub-domains to achieve appropriate size for a development team. External experts were consulted to identify best practices and avoid anti-patterns. Still, services got often merged or split up again, which according to P14, improved the system's maturity in a continuous evolution process.

He also recognized drawbacks of the architecture like low synergies between teams and costly operation. The rather small services also lead to a high complexity at the macro level. Still, the improved time-to-market would compensate for all of these issues. To improve collaboration across teams, the idea of guilds [23] was successfully implemented. In this regard, P14 emphasized the importance of an ongoing coaching by external consultants for technical "*and*" process-related topics. In retrospect, he saw a big threat in taking processes and organizational changes too lightly in such a transition where "*teams were often left alone*" (P14). Despite these challenges, the culture changed towards a more open and unconstrained handling. The established DevOps practices strengthened the team spirit and eased handling duties or absences of people.

**C9-S13, Automotive End-User Services Management System:** Software and IT services company C9 was in the process of migrating a traditional WebSphere system of an external automotive customer which handles management and payment of end-user services in the car. As intentions for a Microservices architecture the customer stated cloud-readiness and shorter, independent release cycles with zero downtime updates. Accordingly, the initial focus was on containerization and hosting the solution in a cloud environment. Service cutting was postponed and then performed applying the strangler pattern [24] for ongoing development. A major technical challenge was removing the application's session state to achieve statelessness [25] and thus enable dynamic scaling.

Mastering technologies took a significant amount of time too, according to P15 around 25% of development efforts. To facilitate distribution of skills and domain knowledge, the cross-functional teams rotate their members from time to time. Decisions by the customer were sometimes politically motivated, like moving from Cloud Foundry to Kubernetes or switching cloud providers. P15 recommended to always be skeptical of technology hypes and criticised the sometimes missing awareness at management level: "*It's for free and* *from Google, so we go with it.*" In his opinion, organizational changes and management decisions would affect people and teams to a greater extent than technical challenges.

**C10-S14, Retail Online Shop:** Similar to the other two online shops (S7, S12), retailer C10 operated a COTS solution (Intershop Enfinity). The system with 200 instances on 100 manually configured blade servers was at its limits technically: "*Response times were disastrous*" (P17). The heavily customized solution made it also difficult to apply regular updates. A glitch of a wrongly priced offering which could not be promptly removed from the shop was the trigger for a modernization. The architects decided to develop a custom in-house solution in a greenfield approach. The architecture was initially based on the concept of Self-Contained Systems (SCS). Over the years, the architecture gradually evolved towards Microservices, as P17 put it in a nutshell: "*The more horizontal aspects you remove and the more experience you have, the smaller the services you can build.*"

Referring to the rich body of available literature, he saw not too many obstacles regarding technologies and development of Microservices, just operating them would become more complex. He considered organizational challenges in a migration process much more crucial. There was a great need for workshops, coaching, and retrospectives to gain the trust of people originally operating the COTS solution: "*A single person can be easily convinced, but a group develops its own dynamics*" (P17). Especially the move from a project-driven to a product-driven development mode was troublesome. Here, the Scrum framework served well in giving orientation to teams and management likewise. P17 emphasized that agility puts people before processes. Accordingly, retrospectives were taken seriously and as a result, most teams shifted from Scrum to Kanban or adapted it to their own needs. In retrospect, the efficiency of the teams and quality of the solution improved dramatically with much shorter release cycles and time to market.

## V. RESULTS AND DISCUSSION

In the following subsections, we present the aggregated results for each research question. Table III summarizes them for all aspects at a glance. In each section (*Intentions*, *Strategies*, *Challenges*), tagging multiple aspects per system was possible. A cross mark indicates the presence of a certain aspect for a system. We considered all aspects named for 2 or more systems and listed them in descending order by number of mentions.

### A. Migration Intentions (RQ1)

The first section of Table III reflects the participants' intentions for migrating their systems. This commonly combines reasons for replacing the legacy system and reasons for choosing a Microservices architecture. Main drivers for replacing the legacy system were a lack of maintainability with different symptoms: overview got lost (S1, S10), changes are too costly (S4) or are prone to cause side effects (S11), and bad analyzability (S11). Furthermore, the investigated systems suffered from operability issues like missing traceability (S11), long startup times (S9), downtime during updates (S6), or difficulties to apply updates in general (S14). P2 referred to their system as having reached the end of its lifetime, it just "*didn't work anymore*". Performance aspects including memory consumption and bandwidth bottlenecks were issues for three systems (S6, S9, S14). Several participants state that current functional requirements cannot be addressed due to deprecated technologies (S6, S11) or cannot be implemented within existing solution by design (S3, S7, S12). Time to market for new features was a driver for several systems (S6, S9, S12, S13), A special case was S6, for which a major change could take up to six months to go into production.

The main drivers for choosing Microservices were scalability of the architecture (S2, S5, S6, S7, S9, S12, S14) as well as for development teams (S1, S3, S6, S7) which for S3 would also facilitate a multi-vendor strategy. Aiming for smaller and better manageable and maintainable units was explicitly mentioned for S4, S5 and S11. In some cases, a company-wide strategy prescribed implementing Microservices (S3, S9, S10), which was not always perceived as justified: "*It was hip and so they went for it*" (P3). Achieving cloud readiness was a driver for two systems (S9, S13), as well as for the quality attributes interoperability (S4, S6) and reliability (S6, S13). As just three out of the 14 systems had finished their migration activities (see Table II), results were only partly obtained to that effect. P1, P7, and P16 confirmed faster decisions and a significantly shorter time to market. The number of defects had decreased for P4, while P2 was confident for the system's long-term maintainability as well.

### B. Migration Strategies and Decomposition (RQ2)

The prevalent migration strategy for our investigated systems was rewriting the existing application (9 cases) combined with extending functionality (7 cases). The *Strangler* pattern [24] was used in seven cases to gradually replace the existing system with Microservices. P2 confirmed that splitting up a complex system would often be too difficult: "*Most productive and successful Microservices projects that I know are greenfield developments.*" Similarly, P7 considered their system as too big for a migration: "*The best consultant on earth can't grasp what they have built over 10 years.*" Systems S5, S7, S12, and S14 started as a Microservices greenfield development, whereas the latter three replaced a COTS solution. In five cases, old and new system were operated in parallel, at times with varying load balancing (S3, S9, S12, S13, S14). Only two systems (S8, S11) were modernized in a continuous evolution process with moderate to low efforts.

The time frames for the investigated migrations range from 1.5 to over 3 years, based on the already finished migrations or expected dates (see Table II). In several cases, a large team was assembled for the initial development and later reduced, e.g. S2 formed teams of up to 40 people to shorten the duration. For some larger projects, several contractors were involved or the system passed through several migration

TABLE III
INTENTIONS, STRATEGIES, AND CHALLENGES PER SYSTEM

| | | S1 | S2 | S3 | S4 | S5 | S6 | S7 | S8 | S9 | S10 | S11 | S12 | S13 | S14 | # of Mentions |
|---|---|---|---|---|---|---|---|---|---|---|---|---|---|---|---|---|
| **Intentions** | Maintainability[a] | x | x | | x | x | x | | x | x | x | x | | | x | 10 |
| | Scalability[b] | | x | | | x | x | x | | x | | | x | | x | 7 |
| | Functional Requirements | | | x | | | x | x | | | x | | x | | | 5 |
| | Operability | | x | | | x | | | | x | | x | | | x | 5 |
| | Company Strategy[c] | | | x | | | | | | x | x | | | x | | 4 |
| | Team Scalability | x | x | | | x | x | | | | | | | | | 4 |
| | Time to Market | | | | | | x | | | x | | | x | x | | 4 |
| | Interoperabiliy | | | | x | | x | | | | | | | | | 2 |
| | Reliability | | | | | | | | x | | | | | x | | 2 |
| **Strategies** | **Process** | S1 | S2 | S3 | S4 | S5 | S6 | S7 | S8 | S9 | S10 | S11 | S12 | S13 | S14 | # of Mentions |
| | Rewrite | x | x | x | x | | x | x | | x | x | | | x | | 9 |
| | Strangler Pattern | x | | x | | x | x | x | | | | x | | x | | 7 |
| | Extension | | x | x | x | | x | x | | | x | | | x | | 7 |
| | Parallel Operation | | x | | | | | | | x | | | x | x | x | 5 |
| | Greenfield | | | | | x | | x | | | | | x | | x | 4 |
| | COTS Replacement | | | | | | | | x | | | | x | | x | 3 |
| | Continuous Evolution | | | | | | | | | x | | | x | | | 2 |
| | **Decomposition** | | | | | | | | | | | | | | | |
| | Other (or non-systematic) | x | x | x | | x | x | | x | x | x | x | | | | 9 |
| | Functional Decomposition | | | | x | | x | x | | | x | | x | x | x | 7 |
| | Existing System's Structure | | x | x | | | | | x | x | | x | | | | 5 |
| | Domain-Driven Design | | | | x | | | | | | | | x | | x | 3 |
| **Challenges** | **Technical** | S1 | S2 | S3 | S4 | S5 | S6 | S7 | S8 | S9 | S10 | S11 | S12 | S13 | S14 | # of Mentions |
| | Decomposition | | x | x | x | | x | | x | x | x | | x | | | 8 |
| | Lack of Expertise | | | x | | x | | | x | x | x | x | x | x | | 8 |
| | DevOps and Automation | | x | | | x | | | x | x | x | x | | | | 6 |
| | Integration of Services | x | x | | | | x | | | | x | | | | | 4 |
| | Legacy System[d] | | | x | | x | | | | | x | | x | | | 4 |
| | Security | | x | | x | x | | | | | | | | x | | 4 |
| | Fault Tolerance | x | | | x | | | | | | | | | | | 2 |
| | **Organizational** | | | | | | | | | | | | | | | |
| | Mindset Change | | x | | x | x | | x | x | x | x | x | x | x | | 9 |
| | Collaboration between Teams | | x | x | | x | | x | x | | | x | x | | | 7 |
| | Justification to Mgmt./Cust. | | x | | x | x | | | | | x | | x | x | | 6 |
| | Recruiting Personnel | x | x | x | | | x | | x | | | | | x | | 6 |
| | Central Governance | | | | x | | | | | x | | | | x | | 3 |
| | Volatile Requirements | | | | | | | x | | | x | | | x | | 3 |

[a] including Analyzability and Modifiability (7), Size and Complexity of Exising Solution (6), Small Units (5)
[b] including Traffic Bottlenecks (3)
[c] including Could Readiness (2)
[d] Complexity, Peculiarities, Deprecated Technologies

project phases until transitioning into a continuous product development mode.

One of the most extensive conversations for P12's team arose from the question: "*What is a service, how big is it and what should it contain?*" Decomposition is the process of splitting up a system or problem space into smaller parts. Seven participants used a functional decomposition approach (S4, S6, S7, S10, S12, S13, S14), as postulated by Microservices design principles [24]. Even though Domain-Driven Design is often cited in literature to achieve such service cuts [5], only three of them reported its explicit usage (S4, S12, S14). For the remaining systems, a different or non-systematic approach was used. Here it was often described as the architect's task or the result of architecture group meetings. For these systems we can observe a significant overlap with using the existing system's structure as a basis (Table III). Refactoring approaches described in academic literature that partly offer tool support [26] were not considered by any of our interviewees. When asked directly, they were not aware of any such tools (P6, P14) or convinced that there is no way to automate it (P2, P12). In case of S6, the very extensive documentation of the legacy system was crucial for steering the decomposition, while P12 made positive experiences with learning from the original creators.

Consequently, we could identify the *Wrong Cuts* anti-pattern [6] in four cases (S3, S4, S9, S12). Two participants commented as follows: "*Our service cut was not good, we sometimes needed to touch two or three services for a change*" (P3), "*The code was not split up appropriately, traffic overhead and data volume increased unnecessarily*" (P11). Furthermore, the *Shared Persistency* (S5) and *Inappropriate Service Intimacy* anti-patterns (S3, S5) were observed as a result of inappropriate cuts. In contrast, other participants appreciated the earned flexibility in merging and splitting services later on (P4, P14, P16), which is less expensive for fine-grained services. However, our participants consciously aimed for more coarse-grained services in a lot of cases. Reasons might be the difficulty in finding the right service cut and a tendency to avoid too complex macro architectures (e.g. S6).

### C. Technical Challenges (RQ3)

Next to finding the right decomposition approach, the lack of expertise in the field of building a Microservices architecture was reported equally often (8 cases). Here, the main issue was to familiarize with the variety of technologies and tools in a timely manner. The often reported difficulties to recruit skilled personnel (P1, P3, P6, P9, P12, P16) reinforced the problem. To fully benefit from Microservices, DevOps practices like build and test automation need to be established. We observed that many organizations faced challenges in this regard or postponed such activities (S2, S5, S8, S9, S10, S11). While most participants reported a decent degree of automation and the usage of CI/CD pipelines, fully automated continuous deployment existed in only 3 cases (S9, S12, S14).

For four systems, integrating the services was a challenge (S1, S2, S7, S11). This also involved interoperability with 3rd-party software or the existing monolith (S11). Size, complexity, or deprecated technologies of the legacy system were obstacles for S3, S6, S11, and S13. Increased effort for assuring the system's security was reported for four systems (S2, S5, S7, S14). For P1 and P5, building a resilient architecture with fault tolerant services was seen as challenging.

However, several interviewees (P4, P5, P8, P15, P16) regarded technical aspects as less challenging, as P8 described it: "*There are always people, maybe external ones, who can master it.*" P4 saw it similarly: "*From 30 years of work experience, I can assure that you get the technology right, but you have to catch people*", which brings us to the second group of challenges.

### D. Organizational Challenges (RQ3)

Our investigated migrations in some cases involved a transition from traditional process models towards agile methodologies. A typical example is S6, where the customer needed to undergo a mindset change from Waterfall to Agile. For architect P7, it was important to cautiously introduce people to the new process and technologies: "*People who realize that their constructed systems and achievements will become obsolete, fear to loose their position in the new world [...] the most difficult thing is to take along everybody and convince them of the new strategy.*" P16 made similar experiences, while P14 emphasized to not take processes and organization too lightly in such a transition. P11 expressed his perceptions from a developer's point of view: "*Politically motivated changes are always difficult, because an organization changes quicker than the software.*" For his team, this situation led to a missing high-level scope, volatile requirements, and incoherent results.

Next to initiating a mindset change towards agile methodologies, collaboration between autonomous teams was seen challenging in seven cases. P6 experienced that the vertical cuts caused pain points in horizontal collaboration. Companies C6, C8, and C10 tried to mitigate the problem by regular agile coaching. For several organizations, the initial months of restructuring caused constant frictions in collaboration (P3, P4, P10).

According to Table II, the vast majority of organizations implemented the Scrum framework. P16 was convinced that it would likely evolve to e.g. Kanban, if Scrum retrospectives are taken seriously. P8 similarly experienced a move towards the *pull principle*. Eventually, several interviewees emphasized the importance of putting people over processes [27] (P7, P8, P11): "*Be it Scrum, Kanban, or even Waterfall: the process doesn't matter with the right people and teams*" (P8). In this regard, P7 appreciated a culture where the ability to experiment and fail was valued.

A migration process involved several activities running in parallel to normal operation. Investments therefore needed to be justified to management or the customer. In six cases, allocating the necessary resources was reported as challenging at times (S3, S5, S6, S11, S13, S14). As Microservices rely on decentralization, central governance and prescribed technological or architectural decisions were sometimes seen as problematic (S4, S9, S13).

## VI. THREATS TO VALIDITY

Similar to other empirical studies, several limitations have to be considered. One such threat is the sampling method, in our case the recruiting of participants. We payed attention to achieve a variance with regards to industry branches as well as company and system sizes. We believe that all interviewees had sufficient expertise and experience in the field. However, not every interviewee had the same level of involvement into all analyzed areas. Our participants were not afraid to also talk about negative sides, albeit in some cases they may not have revealed their true opinion. We tried to mitigate this threat by guaranteed confidentiality and anonymity. Likewise, interviewees may not have remembered all the details or could have misunderstood questions. To address this, we sent the transcripts to each participant for review so that they could correct their answers. To alleviate researcher bias and threats to interpretation validity, all interviews were moderated by the first two authors, who also proofread every transcript. In case of any doubt and to avoid potential misinterpretations, the subsequent research progress was discussed in regular meetings. Albeit we aimed for objectivity in reporting, some parts may be subjective to our interpretation.

With regards to generalizability, we cannot extrapolate from 14 cases to common patterns, e.g. the use of decomposition approaches. Since this is a qualitative study, we focus on participants' rationales and the relations within individual cases. Another limitation could be the geographic location of participants in Germany. Furthermore, the majority of the participants worked for software and IT services companies. As all of those companies migrated systems for different external customers, we are still confident to comprise a diverse sample in terms of sizes, domains, and cultures.

## VII. CONCLUSION

In 16 semi-structured interviews with German-based participants we talked about 14 systems that were in the process of a migration to Microservices or had been migrated recently. Investigated aspects were intentions, strategies, and challenges that participants faced in this process. Similar to related studies [15], [16], maintainability and scalability were identified as main drivers for a migration. Microservices were seen as the appropriate architectural style to address these issues, but were sometimes also prescribed as a company-wide strategy. Functional requirements, i.e. the inability to realize new features with the existing system was a driver in only five cases.

The majority of systems was rewritten using current technologies, while existing code bases were refactored in just two cases. This confirms a tendency discovered by Di Franceso et al. [14]. During the migration phase, old and new system were either operated in parallel, or the *Strangler* pattern was used to successively replace the monolith. For determining the service cut, most companies chose a non-systematic or individual approach. To the same extent, functional decomposition was applied, while the explicit use of Domain-Driven Design was confirmed by only three interviewees. Even though the participants of [14] predominantly used of this concept, decomposition and finding the right service cut were identified as a major challenge in both studies.

As a second major challenge we identified the lack of experience with Microservices-related technologies and concepts, similar to the findings by Knoche et al. [16]. This was reinforced by the difficulty to recruit skilled developers, as reported in six cases. We equally focused on such organizational challenges, which were considered partly more important by our interviewees. While technical difficulties were reported across organizations of all sizes, the larger systems predominantly had issues managing the large number of teams and facilitating the collaboration between them. In particular, mature organizations had to cope with initiating a mindset change when simultaneously establishing agile processes.

Researchers creating industry-focused methods should take these insights into account. Future research could perform deeper analysis of discovered strategies, rationales, and challenges to facilitate the migration of legacy monolithic systems to Microservices.

## ACKNOWLEDGMENT

This research was partially funded by the Ministry of Science of Baden-Württemberg, Germany, for the Doctoral Program "Services Computing" (http://www.services-computing.de/?lang=en).